\documentclass[a4paper]{jpconf}
\usepackage{graphicx}
\usepackage[intlimits]{amsmath}
\usepackage{amssymb}
\usepackage{longtable}
\usepackage{multirow}
\usepackage{setspace}
\usepackage[latin1]{inputenc}
\usepackage{textcomp}         
\usepackage{array}
\usepackage{nicefrac}
\usepackage{slashed}
\usepackage[usenames,table]{xcolor}
\usepackage[colorlinks=true,linkcolor=blue,citecolor=blue,urlcolor=blue]{hyperref}

\begin{document}

\title{Regge behaviour within the Bethe-Salpeter approach}
\author{Stanislav Kubrak, Christian S Fischer and Richard Williams} 
\address{Institut f\"ur Theoretische Physik, Justus-Liebig--Universit\"at Giessen, 35392 Giessen, Germany.}

\begin{abstract}
We present a calculation of the spectrum of light and heavy quark bound states in the rainbow-ladder truncation of Dyson-Schwinger/Bethe-Salpeter equations. By extending the formalism include the case 
of total angular momentum $J=3$, we are able to explore Regge trajectories and make prediction of tensor bound states for light and heavy quarkonia. 
\end{abstract}

\section{Introduction}
The theory of QCD presents many interesting challenges, not least of which that the underlying dynamics described in terms of quarks and gluons is obscurred by the process of hadronization.  In this direction, meson spectroscopy remains an interesting topic especially in light of Belle, Babar and BES, where a wealth of new states have been discovered and warrant investigation.

In this work we focus on the functional approach, namely Dyson--Schwinger equations (DSEs) and Bethe--Salpeter equations (BSEs). By extending the formalism to include mesons with higher angular momentum~\cite{Fischer:2014xha}, we can make predictions for states with $J=3$ in the light and heavy quark sector~\cite{Fischer:2014cfa}. This enables the investigation of Regge trajectories in this approach for the first time.
\section{Framework}\label{sec:framework}
The starting point are the Dyson-Schwinger equations (DSEs), the corresponding solutions of which are the one-particle irreducible Green's functions.
Throughout this paper we employ the rainbow-ladder approach, a widely used truncation that preserves the chiral nature of the pion.
The dressed quark propagator, $S^{-1}(p) = Z_f^{-1}(p^2)( i\slashed{p} + M(p^2))$, with $Z_f(p^2)$ the quark wave function and $M(p^2)$ its mass function, is obtained by solution of its DSE
\begin{align}\label{eqn:quarkdse}
S^{-1}(p)  = Z_2S_0^{-1}(p) 
           + 4\pi Z_2^2 C_F \int \frac{d^4k}{(2\pi)^4} \gamma^\mu S(k+p) \gamma^\nu (\delta_{\mu\nu} - k_\mu k_\nu/k^2)\,  \,\frac{\alpha_{\mathrm{eff}}(k^2)}{k^2}\;,
\end{align}
which we have already specialised to the case of rainbow-ladder by combining the dressed gluon propagator and quark-gluon vertex to form an effective one-gluon exchange model with effective coupling $\alpha_{\mathrm{eff}}(k^2)$~\cite{Bhagwat:2006pu,Blank:2011ha,Hilger:2014nma}.
The bare propagator is $S_0^{-1}(p) = i\slashed{p} + m_0$, with bare quark mass $m_0$ related to the renormalized one via $Z_2 \,m_0=Z_2\,Z_m\, m_q$ and the renormalisation factors $Z_2$, $Z_m$. The $C_F=4/3$ is the usual color Casimir.

For our effective interaction we employ the frequently used model of Maris and Tandy~\cite{Maris:1999nt}. This consists of an ultraviolet part that guarantees agreements with one-loop resummed perturbation theory, and an infrared term whose integrated strength triggers dynamical chiral symmetry breaking. It has the form
\begin{align}\label{eqn:generalmaristandy}
  \alpha_{\mathrm{eff}}(q^2)=\pi\eta^7x^2e^{-\eta^2x} \; + \; \frac{2\pi\gamma_m\left(1-e^{-y}\right)}{\ln\left[e^2-1+(1+z)^2\right]}\;,
\end{align}
with momenta $x=q^2/\Lambda^2$, $y=q^2/\Lambda_t^2$, $z=q^2/\Lambda_{\mathrm{QCD}}^2$.
Typically $\Lambda$ is constrained in the light quark sector to be equal to 
$0.72\,\mbox{GeV}$ in order to set the scale of an non-pertubative interaction such, that the pion mass and
leptonic decay constant are reproduced correctly. In order to incorporate flavor dependence, the $\eta$ parameter differs in the effective 
interaction for light and heavy hadron states. 

Bound states of a quark and an antiquark are described by the Bethe--Salpeter equation
for the corresponding Bethe--Salpeter amplitude $\Gamma(p;P)$
\begin{align}
  \left[\Gamma(p;P)\right] = \lambda(P_i^2) 4\pi Z_2^2 C_F \int \frac{d^4k}{(2\pi)^4}
                                  \gamma^\mu \left[S_+\Gamma(k;P)S_-\right] \gamma^\nu (\delta_{\mu\nu} - k_\mu k_\nu/k^2) \, \frac{\alpha_{\mathrm{eff}}(k^2)}{k^2}\,.
\end{align}
This equation can be seen as an eigenvalue problem, whose solutions $\lambda\left(P_i^2\right)=1$ found at $P^2=-M_i^2$ form a discrete spectrum. The quantum numbers of the meson follow from the covariant structure of $\Gamma(p;P)$.

\begin{figure*}[b]
\begin{center}
\includegraphics[width=0.80\textwidth]{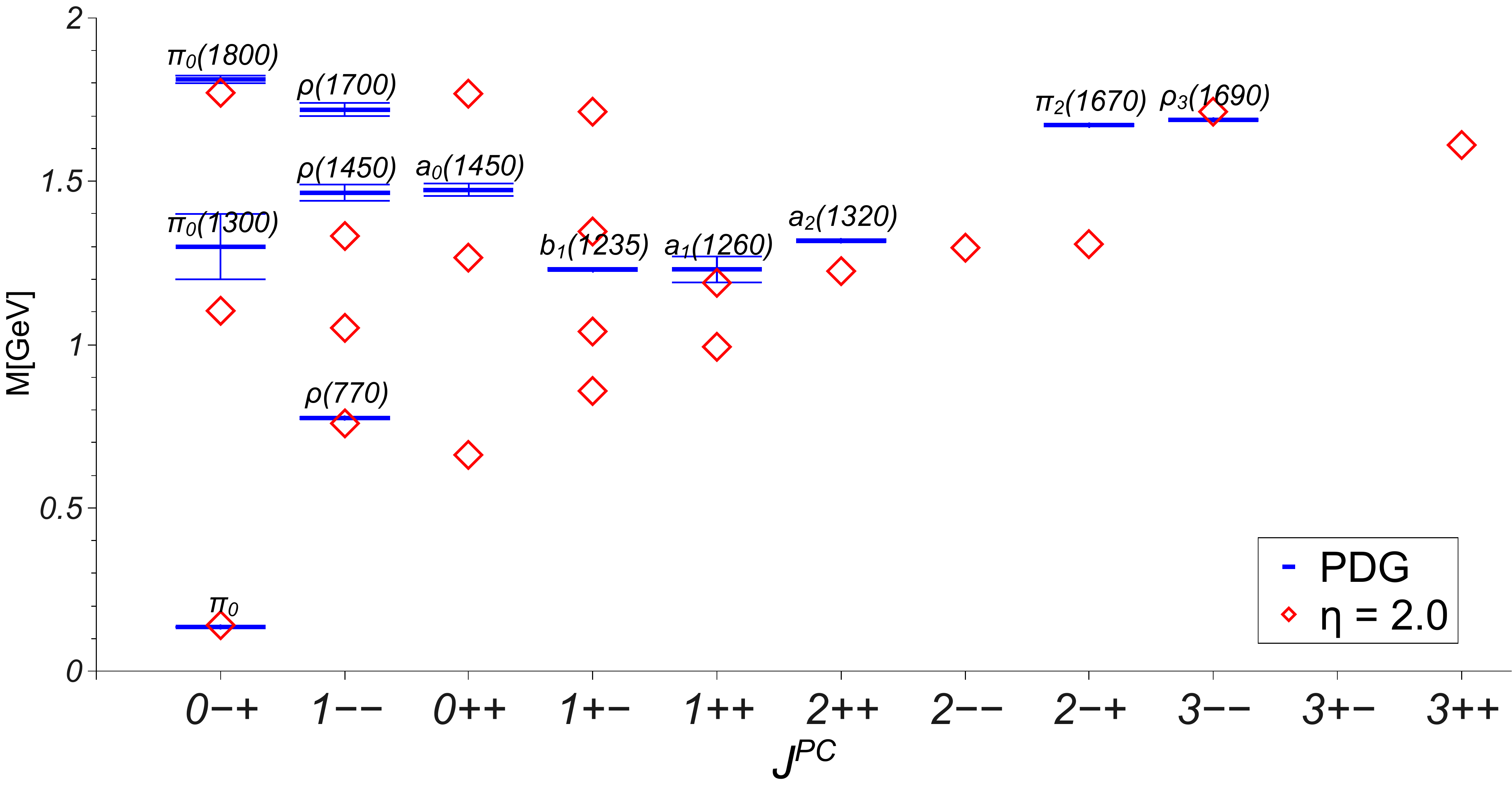}
\caption{\footnotesize The calculated $n\bar{n}$ spectrum, compared to experiment. }\label{fig:spectrumnn}
\end{center}
\end{figure*}

\section{Results}\label{sec:results}
In Fig.~\ref{fig:spectrumnn} we show the resulting spectrum for $n\bar{n}$ mesons.
With the parameters $\Lambda$ and $\eta$ to be adjusted such, that the pion mass is fixed to experimental value, the ground state mass in the vector channel $\rho$ agrees with experimental value. However, further on the spectra, for the scalar and axialvector states we see a significant deviation in masses. Nevertheless, the situation greatly improves for the two tensor channels with $J=2^{++}$ and $J=3^{--}$.

\begin{figure*}[t]
\begin{center}
\includegraphics[width=0.80\textwidth]{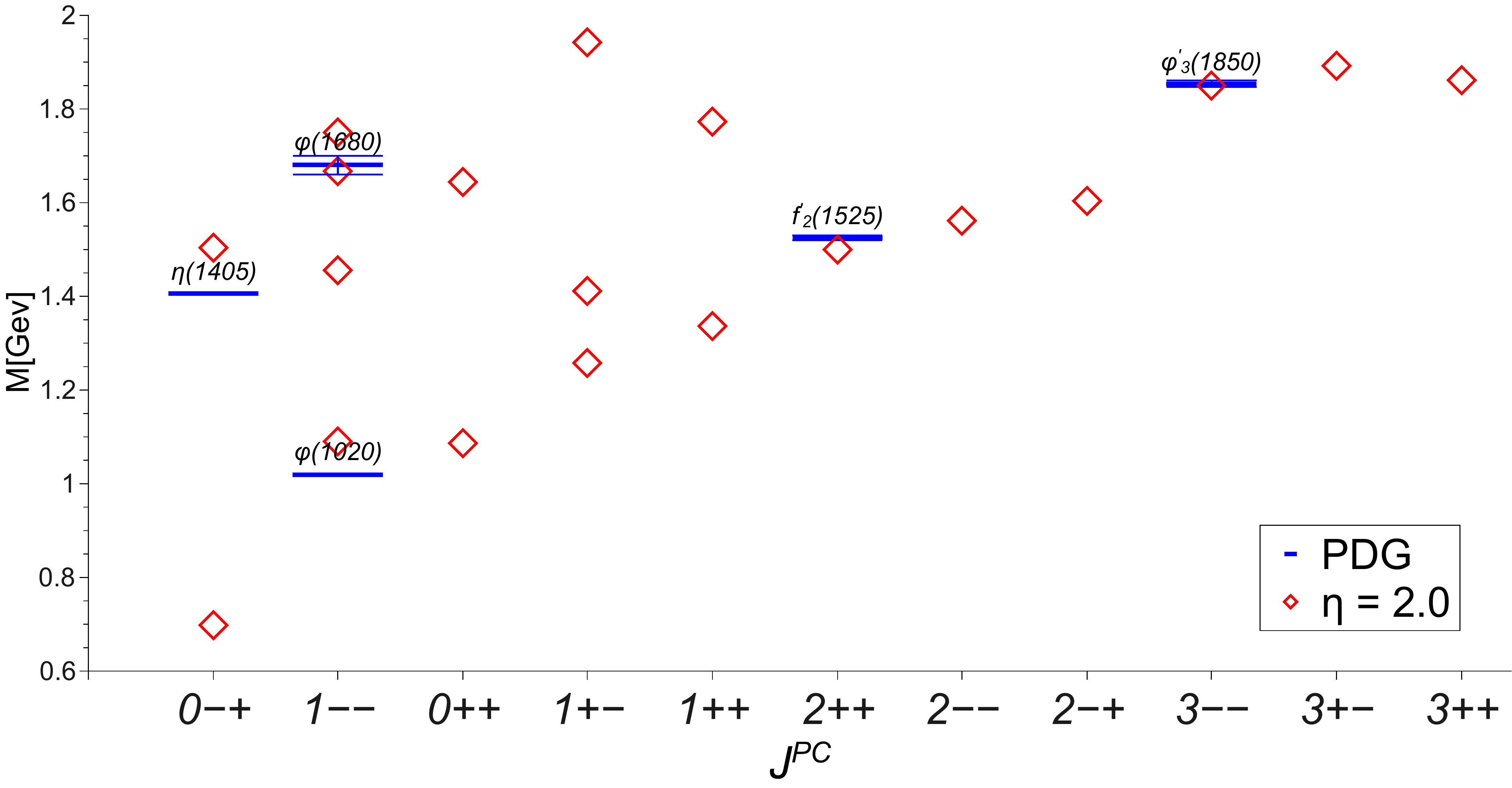}
\caption{\footnotesize Calculated $s\bar{s}$ spectrum, compared to experiment.}\label{fig:spectrumss}
\end{center}
\end{figure*}

It is apparent that the sequence of ground states lying on the Regge trajectory with quantum numbers $J^{PC}=1^{--},2^{++},3^{--}$ is in good agreement with experiment.
The physical interpretation of this is as follows: in the scalar and axial-vector states the tensor force plays a significant role, but for the $J^{PC}=2^{++},3^{--}$ states by contrast it is vanishing or small, whereas the spin-orbit forces are dominant. Translating our observation to a naive quark model picture, we conclude that in the rainbow-ladder truncation the analogue of the tensor force is not sufficient.
The $s\bar{s}$ spectrum is displayed in Fig.~\ref{fig:spectrumss}.
As before one observes a similar picture in that the sequence $J^{PC}=1^{--},2^{++},3^{--}$ agrees fairly well with experiment.
However the pseudoscalar $s\bar{s}$-state is too light because effects from the $U_A(1)$ anomaly~\cite{Alkofer:2008et} are not taken into account. In the $J^{PC}=2^{++}$ and $J^{PC}=3^{--}$ channels, we closely reproduce the experimental values for the $f_2(1525)$ and the $\varphi_3(1850)$. 

\begin{figure*}[b!]
  \begin{center}
  \includegraphics[width=0.80\textwidth]{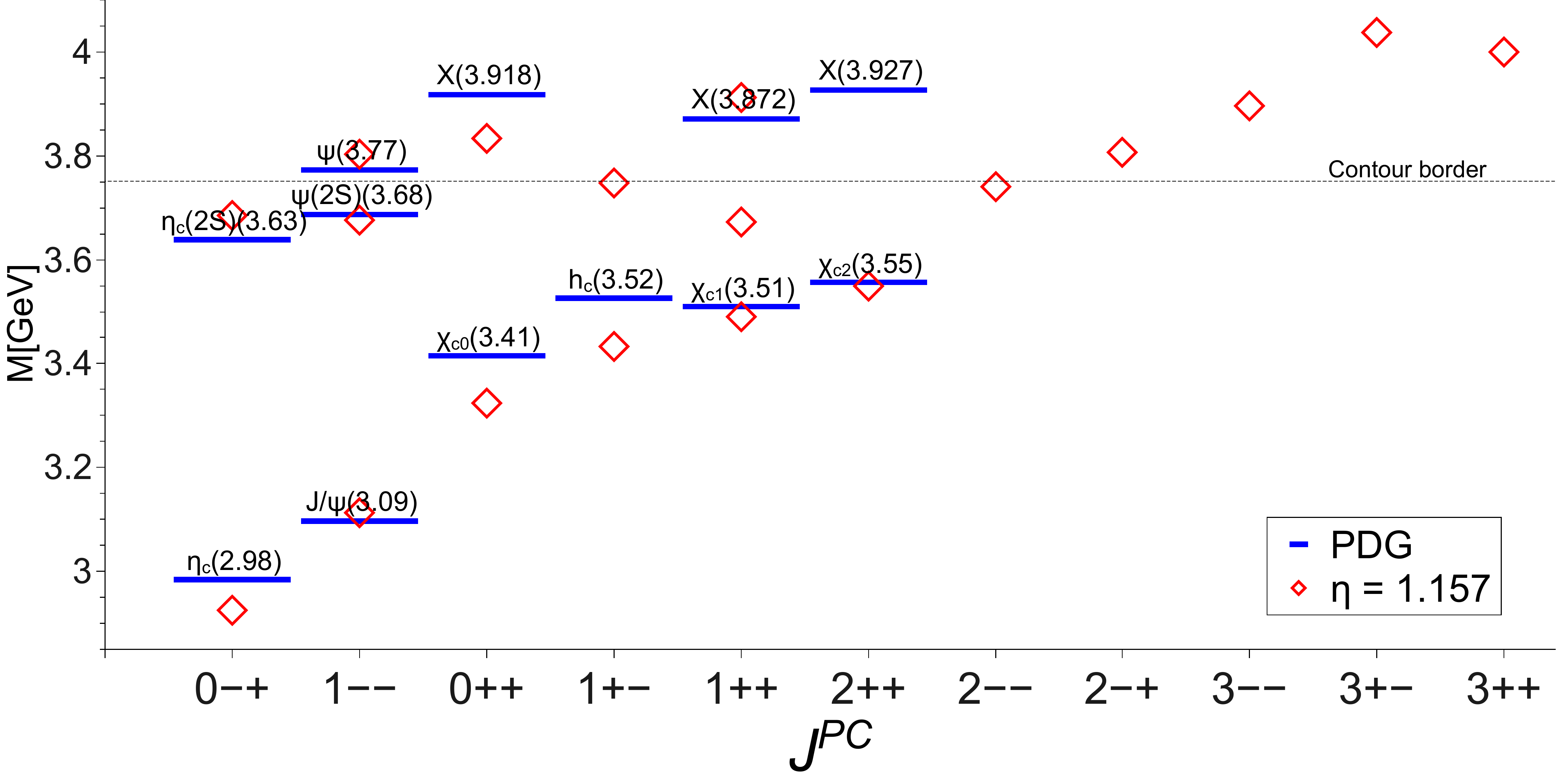}
  \caption{\footnotesize Spectrum of ground and excited charmonium states.}
  \label{fig:charm}
  \end{center}
\end{figure*}

For the charmonium sector the $J/\Psi$ and $X_{c2}$ states masses were used to fix the two model parameters, $\eta$ and the quark mass. The remaining spectrum can be viewed as predictions. 
In addition to the ground states, we observe a good agreement for excited states in the scalar and vector channel. The masses of $\eta_c^\prime$ and $\Psi(2S)$ states coincide with experiment within a $1\%$ percent accuracy. Even the third state in vector channel, the $\Psi(3S)$, is nicely represented. Unfortunately any further excitations are unreachable due to the non-analytic structure of the quark dressing functions at large $M_i^2$, as well as the onset of the $D\bar D$-threshold.

\begin{figure*}[t!]
  \begin{center}
    \includegraphics[width=0.80\textwidth]{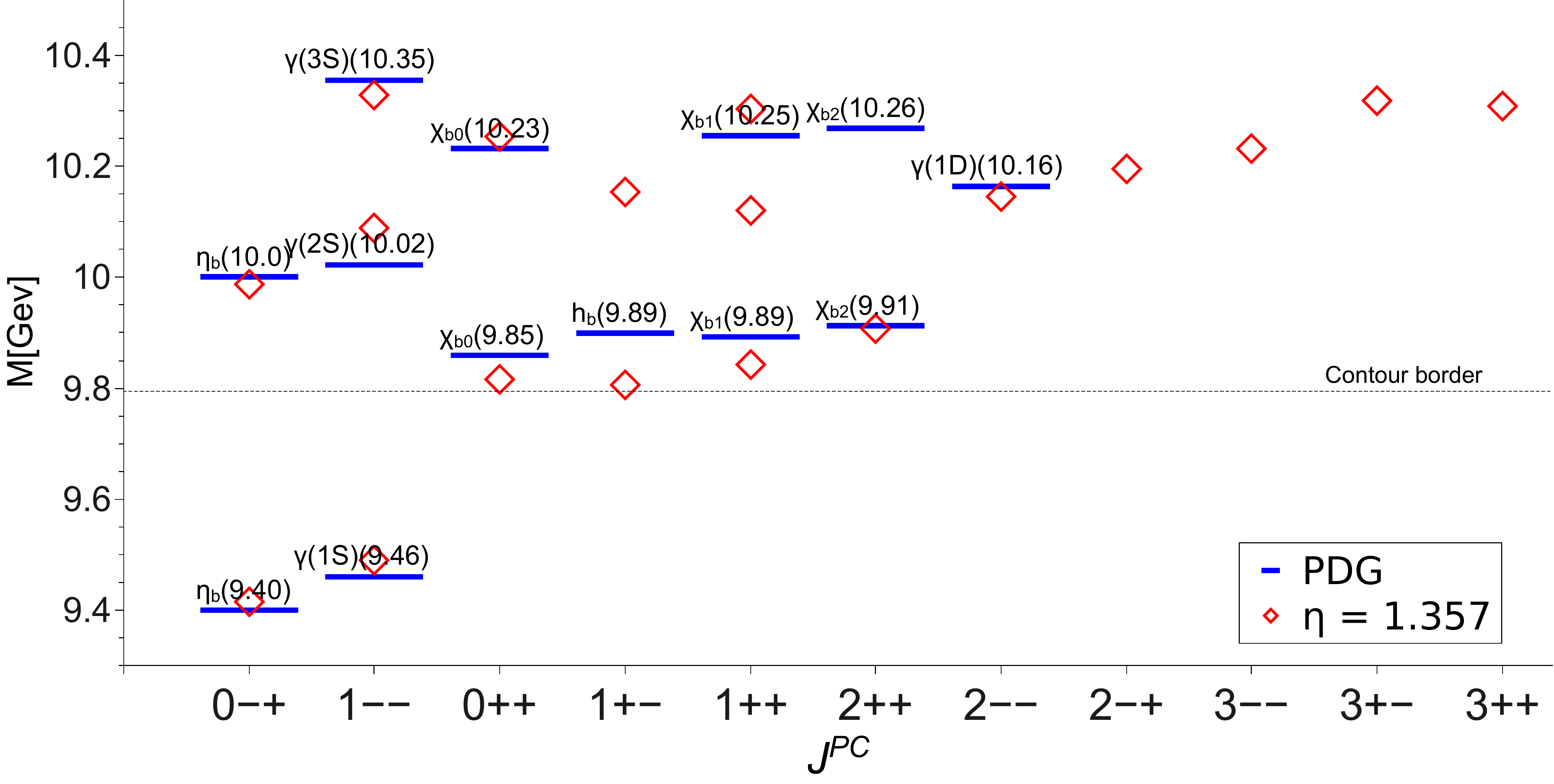}
    \caption{\footnotesize Spectrum of ground and excited bottomonium states.}\label{fig:bottom}
  \end{center}
\end{figure*}

Similarly, for bottomonium the spectrum, shown in Fig.~\ref{fig:bottom}, is also very well represented for the ground-states, agreeing well with experimental data. Note that the $\eta$-parameter was changed to $\eta=1.357$ in comparison to a charmonium calculation, in order to take into account the intrinsic flavour dependence of the quark-gluon interaction~\cite{Williams:2014iea}.

Further, we apply an observation made in light quark sector, that the channels 
$1^{--},2^{++},3^{--}$ (the natural Regge trajectory) are well described by the rainbow-ladder truncation, to predict the charmonium and bottomonium $3^{--}$-state. We find $m^{3^{--}}_{charm}= 3.896 \,\mbox{GeV}$ and $m^{3^{--}}_{bottom}= 10.169 \,\mbox{GeV}$. The result agrees with the quark model prediction~\cite{Ebert:2011jc} and the lattice QCD results \cite{Bali:2011rd,Liu:2012ze}.

Finally, we extract the parameters of the Regge-trajectory given by $ M^2 = M^2(0) + \beta J$, with the slope parameters $M^2(0)$ and $\beta$ given in Table~\ref{tab:regge}.
That we obtain Regge behaviour is intriguing, because the model that we employ cannot (directly) be related to a linear rising quark potential. 
\begin{table}[!th]
\begin{center}
\renewcommand{\arraystretch}{1.3}
\begin{tabular}{c|cccc}
\hline
\hline
          & $n\bar{n}$  &  $s\bar{s}$  &  $c\bar{c}$  &  $b\bar{b}$ \\
\hline
$M^2(0)$  & $-0.42$     &  $0.05$      &  $2.72$      &  $9.12$\\
$\beta$   & $0.99$      &  $1.12$      &  $0.39$      &  $0.371$\\
\hline
\hline
\end{tabular}
\end{center}
\caption{Parameters of Regge trajectories for $n\bar{n}$, $s\bar{s}$, $c\bar{c}$ and $b\bar{b}$ mesons with natural parity. \label{tab:regge}}
\end{table}
\section{Summary and conclusions}\label{sec:conclusions}
We presented a calculation of the spectra of light and heavy mesons within within the 
rainbow-ladder truncation using a well-established effective interaction.

In the light quark sector, the results are reliable on the five percent level for channels where tensor force does not play a dominant role, {\it i.e.} $J^{PC}= 2^{++}, 3^{--}$. 
As a result, we find that the associated Regge trajectory agrees with experiment. Since our approach does
not incorporate a linear rising potential, it is unexpected that we observe Regge trajectories.
This fact may cast some doubts on the naive interpretation of Regge-behavior.

In the case of heavy quarkonia, for spectrum of charmonia and bottomonia we obtained good results, as for ground states as well as for first and second radial excitations in the scalar and vector channels. For the tensor states $3^{--}$ we have enough confidence in our calculation to make a prediction since it lies on the natural Regge trajectory $J^{PC}=1^{--},2^{++},3^{--}$ and is therefore well suited to the rainbow-ladder interaction.

However, the comparison with experiment highlights the necessity to extend this framework beyond that of rainbow-ladder .  
There are a several ways to approaches in this direction: to take into account the effects of the gluon self-interaction
\cite{Fischer:2009jm,HSAWilliams:inprep}, pion cloud \cite{Fischer:2008wy} or through modelling the quark-gluon vertex beyond the leading $\gamma_\mu$~\cite{Chang:2009zb,Chang:2010hb,Chang:2011ei,Heupel:2014ina}.
\section*{Acknowledgments}
We thank the organizers of FAIRNESS2014 conference for giving an opportunity to present the results. 
This work was supported by the BMBF under contract No. 06GI7121, 
the Helmholtz International Center for FAIR within the LOEWE program of the 
State of Hesse.

\vspace{1cm}

\bibliographystyle{iopart-num}

\end{document}